\documentclass{article}

\usepackage{arxiv}

\usepackage[utf8]{inputenc} 
\usepackage[T1]{fontenc}    
\usepackage{hyperref}       
\usepackage{url}            
\usepackage{booktabs}       
\usepackage{amsfonts}       
\usepackage{nicefrac}       
\usepackage{microtype}      
\usepackage{lipsum}
\usepackage{graphicx}
\usepackage{multirow}
\usepackage{float}
\graphicspath{ {./images/} }
\usepackage[dvipsnames]{xcolor}
\usepackage{amsmath}

\title{Ensembles of Convolutional Neural Network models for pediatric pneumonia diagnosis}

\author{
 Helena Liz \\
  Dept. Computer Sciences, Universidad \\
  Rey Juan Carlos\\
  Computer Systems Engineering Department, \\
  Universidad Polit\'{e}cnica de Madrid, Spain\\
  \texttt{helena.lizlopez@gmail.com} \\
   \And
 Manuel S\'{a}nchez-Monta\~{n}\'{e}s \\
  Computer Science Department Universidad \\
  Aut\'{o}noma de Madrid, Spain \\
  \texttt{manuel.smontanes@uam.es} \\
  \And
 Alfredo Tagarro \\
  Pediatrics Department. Hospital \\
  Universitario Infanta Sofía. \\
  Pediatrics Research Group. \\
  Universidad Europea de Madrid, Spain\\
  Pediatric Research and Clinical \\
  Trials Unit (UPIC). \\
  Instituto de Investigación \\
  Sanitaria Hospital 12 de Octubre\\
  (IMAS12), Madrid, Spain. \\
  Fundación para la Investigación \\
  Biomédica del Hospital 12 \\
  de Octubre, Madrid, Spain. \\
  RITIP (Traslational Research \\
  Network in Pediatric Infectious Diseases\\
  \texttt{alfredo.tagarro@salud.madrid.org} \\
   \And
 Sara Dom\'{i}nguez-Rodríguez \\
  Pediatric Research and Clinical \\
  Trials Unit (UPIC). \\
  Instituto de Investigación \\
  Sanitaria Hospital 12 de Octubre\\
  (IMAS12), Madrid, Spain. \\
  Fundación para la Investigación \\
  Biomédica del Hospital 12 \\
  de Octubre, Madrid, Spain. \\
  RITIP (Traslational Research \\
  Network in Pediatric Infectious Diseases\\
  \texttt{sara.dominguez.r@gmail.com} \\
     \And
 Ron Dagan \\
  Faculty of Health Sciences, Ben-Gurion University \\
  of the Negev, Beer-Sheva, Israel\\
  \texttt{rdagan@bgu.ac.il} \\
     \And
 David Camacho \\
  Computer Systems Engineering Department, \\
  Universidad Polit\'{e}cnica de Madrid, Spain\\
  \texttt{david.camacho@upm.es} \\
}

\begin{document}
\maketitle
\begin{abstract}

Pneumonia is a lung infection that causes 15\% of childhood mortality (under 5 years old), \textcolor{Black}{over 800,000 children under five every year, around 2,200 every day, all over the world.} This pathology is mainly caused by viruses or bacteria. X-rays imaging analysis is one of the most used methods for pneumonia diagnosis. These clinical images can be analyzed using machine learning methods such as convolutional neural networks (CNN), which learn to extract \textcolor{Black}{critical} features for the classification. However, the usability of these systems is limited in medicine due to \textcolor{Black}{the} lack of \emph{interpretability}, \textcolor{Black}{because of} these models cannot be used to generate an understandable explanation (from a human-based perspective), about how they have reached those results.
Another problem that \textcolor{Black}{difficults} the impact of this technology is the limited amount of labeled data in many medicine domains.
The main contributions of this work are two fold:
\textcolor{Black}{the first one is} the design of a \textcolor{Black}{new explainable artificial intelligence (XAI) technique based on combining the individual \textit{heatmaps} obtained from each model in the ensemble}.
This allows to overcome the explainability and interpretability problems of the CNN "black boxes", highlighting those areas of the image which are more relevant to generate the classification.
The second one is the development of new ensemble deep learning models to classify chest X-rays that allow highly competitive results using small datasets for training. We tested our ensemble model using a small dataset of \textit{pediatric X-rays} (950 samples of children between one month and 16 years old) with low quality and anatomical variability (which represents one of the biggest challenges addressed in this work). We also tested other strategies such as single CNNs trained from scratch and transfer learning using CheXNet. Our results show that our ensemble model clearly outperforms these strategies obtaining highly competitive results. Finally we confirmed the robustness of our approach using another pneumonia diagnosis dataset (Kermany et al. \cite{kermany2018identifying}).
\end{abstract}


\keywords{Ensembles of Convolutional Neural Networks \and eXplainable Artificial Intelligence \and Heatmaps \and Clinical Decision Support Systems \and Pneumonia \and Pediatrics}

\section{Introduction}

Pneumonia is the most common infectious disease in humans and the leading cause of childhood morbidity refers to having a disease or a symptom of disease, or to the amount of disease within a population) and mortality in the world \cite{prina2015community} that causes inflammation of the alveoli \cite{thompson2016jama}. It especially affects children under 2 years old and elderly above 65. Globally, 15\% of childhood mortality is caused by this disease, over 808,694 children in 2017 \cite{who2020}. This mortality still remains around 800,000 children in 2018 under five every year, or 2,200 children every day, resulting, as UNICEF shows in \cite{unicef2020}, the dramatic number of one child dead  every 39 seconds, which includes over 153,000 newborns. Pneumonia is mainly caused by viruses or bacteria. Most frequent associated viruses are respiratory syncytial virus (RSV), influenza virus and human parainfluenza virus (HPIV) \cite{Bhuiyan}. Most frequent bacteria are \textit{Streptococcus pneumoniae}, \textit{Haemophilus influenzae}, \textit{Streptococcus pyogenes} and \textit{Staphylococcus aureus}, and \textit{Mycoplasma pneumoniae} \cite{Bhuiyan}. \textcolor{Black}{The global prevalence of viral pneumonia origin in children is 14-62\%, being higher in children under two years old and decreasing with age} \cite{van2012associations,rodrigues2018community}. 

Children with bacterial pneumonia should receive antibiotics as soon as possible\textcolor{Black}{, whereas children with viral pneumonia usually only need supportive care}. However, antivirals may have a relevant role in the treatment of viral infections \cite{dong2020discovering}. \textcolor{Black}{When in doubt, and given that bacterial pneumonia is more serious, the symptoms are usually resolved by using empirical antibiotic treatment, which in most cases is unnecessary since a virus is the most frequent cause of community-acquired pneumonia (CAP)} \cite{antibioticresistence}. 

\textcolor{Black}{Pneumonia diagnosis is fundamentally clinical, without reaching an etiological diagnosis most of the time. Due to the high economic cost, and the time it takes to obtain results (days, or even weeks), pediatricians often diagnose based on: laboratory tests, x-rays, and the examination of the patient. X-rays are one of the most important to diagnose CAP} \cite{boersma2006reliability}. Furthermore, it is clear that a high proportion of all pneumonia cases are in fact vial-bacterial co-infections, complicating decisions regarding antibiotic administration \cite{ruuskanen2011viral}. Therefore, pediatricians have to decide empirically whether the child needs antibiotics, \textcolor{Black}{and choose the best treatment with their limited available tools}. As a result, most children receive antibiotics. These results in what is considered over-treatment with antibiotics, leading to the need to narrow the indications by an appropriate discriminative diagnosis  \cite{tomson2014need}. Proxies for typical bacterial pneumonia have been proposed, but the consensus is that no single biomarker alone is enough for diagnosing bacterial pneumonia \cite{bhuiyan2019combination,higdon2017association}. The old paradigm that bacterial pneumonia is associated with a specific radiographic pattern different from the pattern of viral pneumonia is now often criticized, although the radiological pictures of alveolar pneumonia \textcolor{Black}{(also termed lobar pneumonia, or \textit{pneumonia with consilidation})} appear to be bacterial in most of the cases \cite{world2001standardization}. The interpretation of the chest X-ray radiography (CXR) is usually performed following the standards of the “WHO Vaccine Trial Investigators Radiology Working Group” \cite{cherian2005standardized}. These standards establish two possible interpretations: “consolidation” (including consolidation and/or pleural effusion as per WHO standards) and “other infiltrates”. However, the inter-observer agreement for these two categories is low \cite{mackenzie2016definition}.

Convolutional Neural Networks (CNNs) are well-known Deep Learning architectures. \textcolor{Black}{Recently} great advances have undergone helping to solve several visual-related tasks \cite{martin2020optimising}. This kind of neural networks is inspired by the biological neurons of the visual cortex \cite{lecun2015deep}, which allows them \textcolor{Black}{to} solve problems such as image classification \cite{he2016deep} and object recognition \cite{ji20123d}, \textcolor{Black}{among  other image and video processing and recognition tasks}. CNNs are also successful in other problems such as speech recognition \cite{hinton2012deep}, malware detection \cite{martin2017evolving,martin2018evodeep}, natural language processing \cite{cambria2014jumping,young2018recent}, among many others. These systems process the information in two main steps: \emph{feature extractor}, where relevant features are detected; and \emph{classification}, where these features obtained from the previous step are analyzed and different probabilities will be assigned to the detected structures to carry out the classification. In areas such as medicine, where the diagnosis is often based on the analysis of clinical images (e.g. x-rays), CNNs and Deep Learning methods have proven both their usefulness and effectiveness in the detection and classification of multiple diseases  \cite{lecun2015deep,litjens2016deep,sirazitdinov2019deep}. \textcolor{Black}{The classification performance in computer vision problems, such as those mentioned above, can be improved through ensembles of models \cite{li2015convolutional,simonyan2014very}. An ensemble is a kind of method based on the combination of multiple learning algorithms, the ensemble is designed to improve the performance of the particular elements that builds up the new algorithm. This technique combine the individual predictions to produce a consensus prediction. Its advantages over individual models are the performance, because the combination of multiples models can improve their individual power and the final model can approximate better the optimal solution \cite{dietterich2000ensemble}; and robustness, because the ensemble models reduce the variance of prediction errors made but the contributing models by adding bias, avoiding overfitting of the final model \cite{minetto2019hydra}. For this reason this technique can be critical in small datasets like ours, where the information is particular limited in both size and quality.}

\textcolor{Black}{However, CNNs, like many other Deep Learning and machine learning methods, are considered as "\emph{black-box}" algorithms, where both the input and output can be easily analysed and interpreted by the users, but where the inference process carried out by the algorithm is opaque hinders end-users' confidence in the results obtained, and therefore makes decision-making negatively affected}. It makes this essential process (”how” and ”why” the algorithm has obtained this outcome) uninterpretable for the human being \cite{guidotti2018survey}. This may limit its application in fields such as medicine, where the practitioners need to know how the algorithm has inferred the output for each specific patient \cite{holzinger2017we} (e.g. why the algorithm is assigning a 90 \% probability for alveolar pneumonia?). This limitation can be overcome using automatic explanatory systems, called explainable AI (XAI) \cite{arrieta2020explainable}, \textcolor{Black}{which allows us to visualize which areas of the image (features) have been used to obtain the outcome generated by the algorithm}, or at least alleviating, the aforementioned problem. These XAI-based systems will generate new images highlighting the areas of highest interest that the system uses to obtain the result (e.g. in our case to predict a particular kind of disease) \cite{fong2017interpretable}.

The combination of Deep Learning models with medical knowledge allows the development of new \textcolor{Black}{clinical decision support systems (CDSS)}.
These automatic systems can help in medical diagnosis reducing some typical clinical problems such as subjectivity in the interpretation of medical tests or human errors (fatigue, distraction, etc.) \cite{mahomed2020computer}.

\textcolor{Black}{This combination of machine learning methods with human-based knowledge can improve the performance of the diagnosis process}, as it was stated in \cite{kontzer}. In that project, leaded by Dr. Andrew Beck, it was demonstrated that the combination of pathologists and Deep Learning models provide a significant reduction \textcolor{Black}{of} the error rate for breast cancer diagnosis. In the initial results pathologists obtained a 3.5\% of error during classification of the pathology, whereas the Deep Learning algorithm obtained a slightly better result of 2.9\% error. However, when both humans and AI model were combined this error decreased to an impressive 0.5 \% (so, the 99.5\% of cases were correctly classified) \cite{kontzer}.

\textcolor{Black}{The main contribution of this work can be briefly summarized as the design and development of a \textcolor{Black}{\emph{novel Machine Learning system based on ensembles}} for pneumonia diagnosis in childhood.
Our \textcolor{Black}{system} estimates the probability that the X-ray has a consolidation or other infiltrates, which would be a helpful tool for the unclear cases in case of disagreement among professionals, or in case of work overload. The result generated from the \textcolor{Black}{system} should be user-friendly (from a health professional perspective), to achieve that, the system creates a graphical visualization using an explainable AI technique named heatmap}, which highlights the areas of the image which are more relevant for the diagnosis according to the AI system \cite{samek2017explainable}. 

An interesting point of this work is to understand how the neural network infers the pneumonia type (alveolar versus non alveolar) from the chest X-ray. This could help to expedite the treatment of patients who require medication. On the other hand, this could also help to avoid giving antibiotics to patients who do not need them. \textcolor{Black}{This is crucial since an incorrect use of antibiotics (that is, using them in patients who do not have a bacterial infection), or an excessive use of broad-spectrum antibiotics, can cause \textbf{antibiotic resistance}}. This can become a global problem making it more difficult to treat patients, the only solution being the development of new, more powerful antibiotics \cite{tomson2014need}. Therefore, in order to reduce the overuse of antibiotics in viral pneumonia, a correct diagnosis of bacterial pneumonia is crucial.

This article has been structured as follows: Section \ref{related_work} provides a short description of some relevant works in the area of AI-based detection of lung diseases; Section \ref{metodology} describes the methodology followed to design and train our CNN models; Section \ref{experimental_results} shows the experimental results; Section \ref{discusions_conclusions} presents the conclusions and finally, future work, with some future lines of work.

\section{Related work}
\label{related_work}

\textcolor{Black}{AI techniques have been intensively applied in medicine, especially in diagnosis processes, forecasting and prediction of diseases evolution, global health impact, etc. We can find a very large number of examples, such as the diagnosis of brain abnormality with magnetic Magnetic resonance imaging (MRI) \cite{talo2019application}, cardiac disorders using clinical data \cite{chang2021usefulness}, breast lesions using radiographs (X-rays) \cite{al2020evaluation} or COVID-19 with chest X-ray \cite{ismael2021deep}. Within the field of pneumonia there is also a multitude of systems with different kinds of data, such as: clinical data \cite{dai2021establishing}, ultrasounds \cite{singh2021classification}, computed tomography \cite{li2020artificial} or X-rays \cite{kermany2018identifying,yue2020comparison}, among many others. However, there are also numerous studies focused on other aspects, such as predicting the evolution of patients \cite{mushtaq2020initial} or classification of patients according to severity
using chest X-rays \cite{tabik2020covidgr,zhu2020deep}. Due to the main contributions of this work, only some relevant works in the area of Deep Learning, and its application to the pathology considered (pneumonia), will be briefly described. }

\subsection{Deep Learning methods for pneumonia diagnosis}
\label{related_work_evolution_techniques}

Most of the previous works in this field use a pediatric dataset published by Kermany et al. \cite{kermany2018identifying}, consisting of a total of 5856 radiographs divided in three classes (2780 bacterial, 1493 viral and 1583 normal) of patients between 1 and 5 years of age. Other works in the field \cite{mahomed2020computer} use other datasets and a classification hierarchy. First, radiographs are classified as "pneumonia" or "normal". Then, radiographs labeled as pneumonia are classified as "viral" or "bacterial". Kermany et al. obtained an AUC of 0.85 for the first classification (pneumonia versus normal) and 0.81 for the second (viral versus bacterial). We can also find the same classification system in \cite{kermany2018identifying}, which will be explained later.

The first step that must be carried out is the pre-processing of the images. There are different algorithms, \textcolor{Black}{depending on the basic features, and pre-processing needs, to apply the target algorithms, of each dataset}. The most commons are Histogram equalization, that enhances pathological signs \cite{sharma2017detection} \cite{khobragade2016automatic}; Lung segmentation, which removes irrelevant data on X-rays and recovers useful information (because consolidation signs only appear inside lungs, therefore the rest of the image is irrelevant for the models). This one is applied in multiplicity of models \cite{li2019attention,mahomed2020computer,kim2021automatic}. 

Currently, one of the most successful AI methods for automatic image classification is Convolutional Neural Networks (CNN) \cite{stephen2019efficient}. Two main problems in CNN training for pneumonia diagnosis (and in most of the medical diagnosis problems) are low dataset availability and small dataset size. To solve these problems several works use \textit{Transfer Learning}, a technique that takes advantage of the knowledge obtained from models used in other (similar) areas and trained with bigger datasets\textcolor{Black}{. T}he idea is that part of the CNN (the first layers) is "inherited" from a model previously trained in other dataset, while the rest of the CNN is trained with the penumonia dataset. This technique was used in the work by Kermany et al. \cite{kermany2018identifying}, where part of the CNN was trained with ImageNet dataset and the rest of the CNN was trained with medical images including pediatric pneumonia. This resulted in an AUC of 0.96 for the classification task "pneumonia" versus "normal", and an AUC of 0.94 for the classification task "viral pneumonia" versus "bacterial pneumonia". This classification also shows an interesting classification distinction, unlike the rest of the networks, between three different categories (normal, bacterial and viral), achieving an AUC of 0.918. Other approaches, such as CheXNet \cite{rajpurkar2017chexnet}, use a CNN called "DenseNet" which was also trained using ImageNet dataset and re-trained with a dataset of 14 different lung diseases (including pneumonia). This model has an AUC value for pneumonia of 0.768, however, the AUC of cardiomegaly and emphysema has higher values than the pneumonia AUC (0.925 and 0.937 respectively). Other technique to solve this problem is \textcolor{Black}{\textit{Data augmentation}: a data-space solution to the problem of limited data, a quiet common problem in this field. It encompasses a suite of techniques that enhance the size and quality of training datasets such that better models can be built using them \cite{elgendi2021effectiveness}, increasing artificially, synthetically, the amount of training data using information only in it, so it helps to avoid overfitting. There are several techniques, such as geometric transformation (flippling, cropping, rotation, translation), color space transformation or noise injection \cite{shorten2019survey}}.

Finally, CNN \textit{ensembles} can improve the performance of a particular model based on an unique architecture, this technique is based on the idea of combining predictions from multiple statistical models (e.g. CNN) to form one final prediction (i.e. a new model
made by a combination of selected models). There exist different ways to generate those ensembles \cite{schwenker2013ensemble}: \textit{averaging}, ensemble averaging creates a group of models, each with low bias and high variance, \textcolor{Black}{then combines them into a new model with (hopefully) low bias and low variance}. Some relevant examples can be seen in Shin et al. \cite{7404017}, which classifies two different datasets: the first one, Thoracoabdominal Lymph Node and Interstitial Lung Disease, comparing the performance of different architectures from the state of the art (CifarNet, Google-Net, AlexNet, etc.), and defining a new architecture that is made by the average value from the considered SoTA methods, or in the work by Christodoulidis et al. \cite{christodoulidis2016multisource}, which designs a system that classify between seven different Interstitial lung diseases using CT, and applying transfer learning technique; \textit{majority voting}, in this approach the new model is built taking into account the outcomes from the different models considered, \textcolor{Black}{there are two approaches to the majority vote prediction for classification;
\textit{hard voting} and \textit{soft voting}}. Hard voting involves summing the predictions for each class label and predicting the class label with the most votes. Soft voting involves summing the predicted probabilities (or probability-like scores) for each class label and predicting the class label with the largest probability. Some examples of majority voting ensembles can be found in Yan et al. \cite{yan2016classification}, which designs a CAD system for lung nodule malignancy risk classification from CT, using different CNNs with the objective of learning different levels of image spatial context, and improving detection performance; and finally, \textcolor{Black}{another popular ensemble method is the \textit{weighted averaging}, which is based on a weighted combination of the former method, where a particular weight will be given to each model before the final model is generated}. Some approches that follows this method as is shown in Bermejo-Peláez et al. \cite{bermejo2020classification}, which designs a model to classify computed tomography between 8 classes within Interstitial Lung Disease, another example of this approach can be found in Sirazitdinov et al. \cite{sirazitdinov2019deep}, which generates an object detection model for pneumonia detection and location from chest X-rays.

\subsection{Explainable AI methods in Medicine}
\label{relatedwork_explainableAI}

\textcolor{Black}{As was previously mentioned, CNNs are considered {\it black-box} algorithms, which increases the difficulty of applying them in areas such as Medicine}. The reason is that even these algorithms can be very precise classifying radiographs or making predictions, the end-users (medical staff in our case) will need to interpret and understand how and why the algorithm has reached the conclusion (\textcolor{Black}{the outcome provided to the end-user). For this reason, it is important to develop eXplainable AI (XAI) systems that allow end-users to understand how the system works (i.e. how and why classifies the radiography in our case)}.

We can distinguish between two main techniques, the first being \textit{object detection} systems. These systems are based on CNN models that locate different objects in images. An example of this type of model is CoupleNet, which classifies and locates signs of pneumonia on a chest radiography and produces a visualization of the original image with bounding boxes in lung areas that show signs of disease \cite{team2018pneumonia}. These algorithms are used in different works to detect and locate pneumonia signs \cite{sirazitdinov2019deep} combining two CNNs for the detection and location of pneumonia signs on lungs.

The second kind of methods uses additional techniques to \textit{visualize how the model classifies}, such as \textit{heatmap} generation. This method is used in a wide variety of problems including pneumonia diagnosis. For example in \cite{mahomed2020computer}, where different lung areas can be seen with different colour intensities according their relevance to the prediction made by the model. Other similar work is presented in Zech et al. \cite{zech2018variable}.

\textcolor{Black}{A substantial difference between the two previous methods is that in the first one it is necessary to build a training dataset where the areas with signs of the disease have been marked by the experts. Whereas, in the second type of methods, it is only necessary to label each radiography of the training set with the classification given by the experts (consolidation / non consolidation in our case)}. After training, medical staff need to review the heatmaps generated by the model \textcolor{Black}{to validate the clinical sense.}
\section{Methodology}
\label{metodology}

\begin{center}
\begin{figure}[h]
    \mbox{\includegraphics[width=6.50in, scale=1]{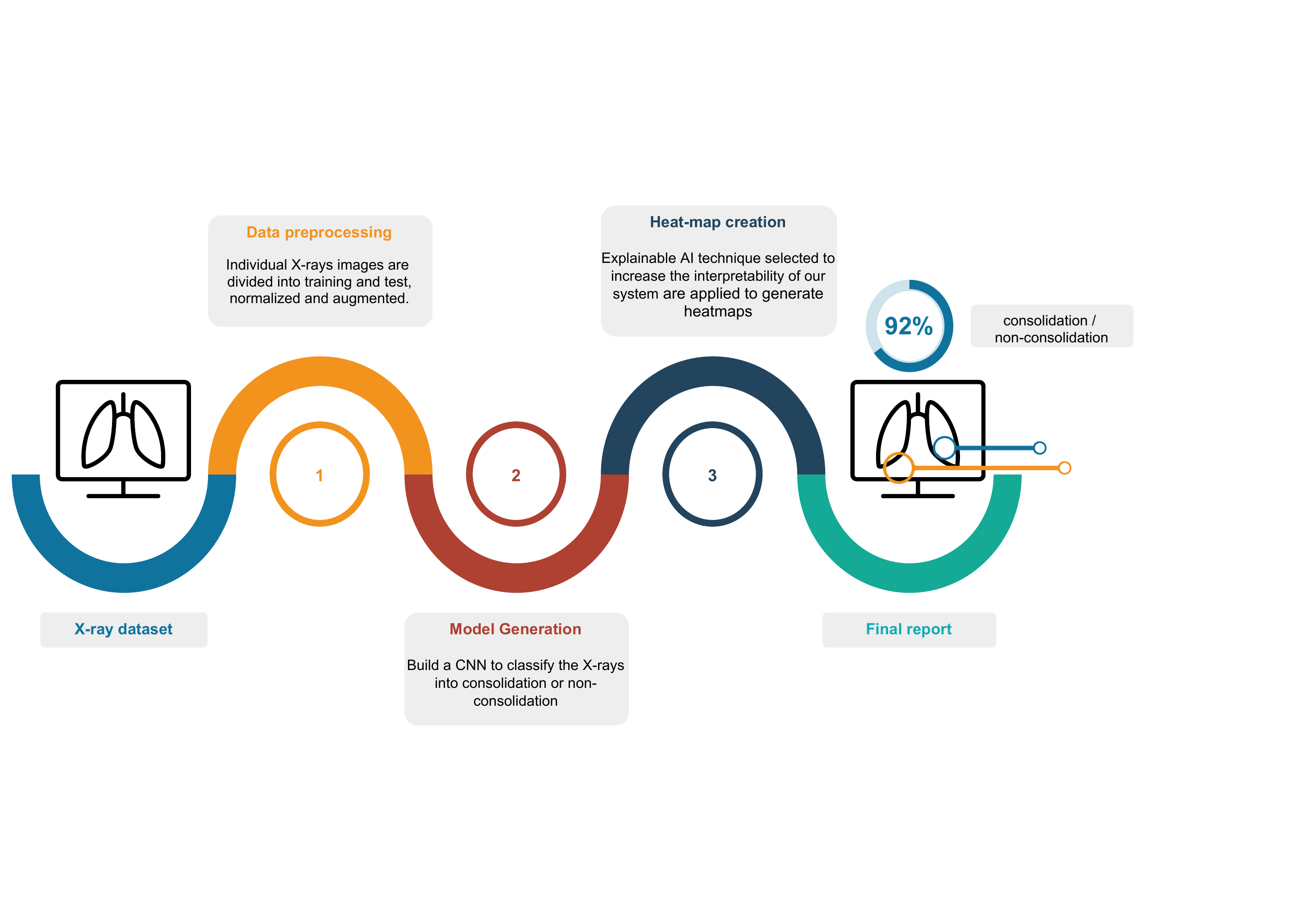}}
  \caption{Data Flow Diagram.}
  \label{fig:software_architecture}
\end{figure}
\end{center}

This section describes the methodology carried out to design and develop our \textcolor{Black}{model} for pneumonia diagnosis.

The system has been designed following the three basic stages shown in Figure \ref{fig:software_architecture}.

All the processing was implemented in Python using well-known libraries such as numpy (mathematical computing) \cite{oliphant2006guide}, matplotlib (visualization) \cite{hunter2007matplotlib}, Keras (Deep Learning, \cite{chollet2015keras}), and Keras-Vis (heatmaps calculation) \cite{raghakotkerasvis}.

The first step is the \textit{data preprocessing} stage where the X-rays dataset is divided into training and test subsets. In this stage, individual X-rays images are also normalized. Data augmentation of training images was performed for a robust model construction \cite{taylor2017improving}.
The second stage is the \textit{generation of a model} that classifies each X-rays into two classes (consolidation / non-consolidation). \textcolor{Black}{The accuracy of this model is validated using a k-fold cross-validation technique (where k, the number of folds, has been fixed to 5).
Finally, the last step is the application of the explainable AI technique that we have selected to increase the interpretability of our} \textcolor{Black}{system}, \textit{heatmap creation} \cite{adadi2018peeking}.
To evaluate the quality of our system we follow two different strategies: 1) generating the heatmap using only a model, and 2) generating the heatmap using an ensemble of models with the same architecture, but trained with different data folds. The second strategy will allow us to compute an uncertainty level (given by the standard deviation) associated with each pixel, \textcolor{Black}{that} allow us to analyze the robustness of the heatmap.


\subsection{Datasets}
\label{met_dataset}

In this work we use two different datasets. The first one is an X-ray pediatric-pneumonia (XrPP) dataset provided by Ben-Gurion University (Israel), and the second one is a public pediatric dataset of chest X-rays \cite{kermany2018identifying}, which will be used to analyze the generalization capability of our model. \textcolor{Black}{X-rays have been labeled by experts using one of the following two mutually exclusive classes}:

\begin{itemize}
\item \textbf{Consolidation}, denoting a chest x-ray image with signs of consolidation (alveolar pneumonia).
\item \textbf{Non-consolidation}, denoting a chest x-ray image with other infiltrates signs that correspond with non-alveolar pneumonia.
\end{itemize}

The first dataset is formed by \textcolor{Black}{950 labeled chest X-rays of children} (between one month and 16 years), 403 cases of consolidation (42.42\%) and 547 cases of non-consolidation (57.58\%). These chest X-rays are posteroanterior (PA) radiographs showing the posterior view of the chest.
Each case was classified by a panel of experts consisting of two senior pediatricians and a radiologist from Hospital 12 de Octubre Research Institute (Madrid, Spain).
The experts also had access to lateral radiographs of the patients to increase the precision of the labeling of each case.
During this classification, 50 samples were withdrawn due to lack of consensus from the expert panel on the \textcolor{Black}{diagnosis\footnote{The dataset is not public and cannot be shared with the community.}.} The authors have obtained the necessary permission from the ethical boards of Ben-Gurion University and Hospital 12 de Octubre Research Institute to work with these data.

Note that the distribution of classes in the dataset is relatively balanced. This is important, especially in small datasets like this, since an unbalanced distribution of classes can severely affect the model's performance (e.g. example accuracy, true positive rate: TPR, false positive rate: FPR) \cite{wei2013}. 

Another interesting feature of the dataset is the size of the images. The average size of the images is similar in both classes (approximately 200,000 pixels), a very low value for X-ray images, which \textcolor{Black}{implies} that image resolution \textcolor{Black}{(and therefore quality)} is limited.
\textcolor{Black}{Therefore, another contribution of our work is the study of the reliability of CNNs when trained with small and low-resolution datasets.}


The second dataset studied in our work is composed by 5,856 X-rays of children between one and five years of age. There are 2,780 cases of bacterial (47.5 \%), 1,493 cases of viral (25.5 \%) and 1,583 cases of normal (27.0\%) \cite{kermany2018identifying}, so the class distribution is not balanced. As described in that paper, all chest radiographs were screened for quality control, removing low quality or unreadable scans. Then, diagnoses for the images were graded by two expert physicians. Since the sizes of the images are between 1,000,000 and 2,000,000 pixels, \textcolor{Black}{both the images resolution (quality) and the number of images are clearly superior to the first dataset}.

\subsection{Data preprocessing}
\label{subsec:met_data_preprocessing}

\textcolor{Black}{The X-ray images were provided by medical centers in jpg format}. This format codes the colour at each pixel using three values, the "RGB" components. In our dataset these values are redundant since RGB components are identical in greyscale images. Therefore we keep only the first one. The original images do not have the same size, so we normalize their shapes to 150x150 pixels. On the other hand, the pixel values of each image are normalized by dividing them by the average pixel value of the image.

As we mentioned before, CNN usually needs a very large number of training images to avoid overfitting, however, our available dataset is particularly small. To overcome this problem we used a popular technique named \emph{Data Augmentation}, which allows us to increase the size of our dataset\cite{taylor2017improving}. It generates batches of images with real-time data augmentation. During each epoch, a different set of variations of the original training images is generated using different types of transformations \cite{chollet2015keras}. In this work shearing (0.2), zoom (0.05), rotation (0.2), horizontal shift (0.1), vertical shift (0.1) and horizontal flip transformations have been used with a batch size of 32.

\subsection{Data partitioning}

In order to robustly evaluate the different architectures for our models, we generated a pool of different training / validation / test partitions of the dataset. For each of those partitions, we first divided the dataset randomly into construction (70\%) and test (30\%) subsets using stratified partitioning. \textcolor{Black}{Then the construction subset was randomly divided into training (80\%) and validation (20\%) subsets using stratified partitioning. Therefore each partition is a division of the original dataset into training (56\%), validation (14\%) and test (30\%) subsets}. These are mutually exclusive, so each image in each partition is only included in one subset (training, validation or test). Training subset will be used to learn the CNN's weights, whereas validation subset will be used for monitoring CNN's metrics throughout model's learning and avoid overfitting. \textcolor{Black}{Finally, test subset have been used for estimating the model's generalization capabilities (performance when new radiographs are considered)}.

\subsection{Convolutional Neural Network model}
\label{met_network_architecture}

\textcolor{Black}{We considered different architectures for our CNN model. The number of convolutional layers in them was in a 3-4 range (see Table \ref{tab:cnn_parameters}), because a low number of convolutional layers may not be able to extract all the information from the images and a high number can lead to overfitting}. Each convolutional layer has 32 kernels and a ReLU activation function. The output of the last convolutional layer is flattened and then perturbed by a Dropout with a rate of 70\%. Then this information is processed by a dense layer ("FC layer") with a number of neurons depending on the architecture (Table \ref{tab:cnn_parameters}) and ReLU activation function. \textcolor{Black}{Finally, the classification layer of our CNN consists in a dense layer of two neurons with Softmax activation}. Therefore, we consider a total of six architectures (Table \ref{tab:cnn_parameters}). Kernel L2 regularization with a strength of 0.01 was applied to FC dense layer. Each CNN was trained using Adam optimizer a learning rate of $1e^{-4}$.

\begin{table}[!h]
\centering
\begin{tabular}{c|c|c|c|c|c|c|}
\cline{2-7}
\multirow{2}{*}{}                                                                                & \multicolumn{6}{c|}{Architectures}                                                                                                                     \\ \cline{2-7} 
                                                                                                 & Arch1 & \multicolumn{1}{l|}{Arch2} & \multicolumn{1}{l|}{Arch3} & \multicolumn{1}{l|}{Arch4} & \multicolumn{1}{l|}{Arch5} & \multicolumn{1}{l|}{Arch6} \\ \hline
\multicolumn{1}{|c|}{\begin{tabular}[c]{@{}c@{}}Number of\\ convolutional layers\end{tabular}}   & 4     & 4                          & 4                          & 3                          & 3                          & 3                          \\ \hline
\multicolumn{1}{|c|}{\begin{tabular}[c]{@{}c@{}}Number of neurons\\ (FC layer)\end{tabular}}     & 64    & 128                        & 256                        & 64                         & 128                        & 256                        \\ \hline
\end{tabular}
l\caption{Hyperparameters of each of the six architectures considered.}
\label{tab:cnn_parameters}
\end{table}

\subsection{Ensemble model}

In order to increase the performance and robustness of our system, ensembles are considered.
Each ensemble is composed by five different CNNs, each constructed using a different partition with different training/validation subsets but same test subset. This ensures models diversity, which is crucial for the ensemble performance \cite{dietterich2000ensemble}.

The partitions were created as follows: first we randomly divided the dataset into construction (70\%) and test (30\%) subsets. Then we generated five different training / validation random partitions of the construction subset (80\% / 20\%). For each of these five partitions a CNN was trained from scratch. The predictions for test subset of the ensemble formed by these five CNNs were then computed and the performance metrics evaluated. Ensemble's prediction for the probability of consolidation (and non consolidation) was computed as the average probability prediction across the five CNNs in the ensemble.

\textcolor{Black}{Summarizing, in order to compute in a robust and consistent way the metrics for the ensemble, the entire process described in this subsection was repeated using five different construction / test divisions. On the other hand, the total number of CNN models built for each division was 5. Therefore, a total of 25 models were constructed}.

\subsection{Reference model}
\label{subsection:reference-model}

\textcolor{Black}{In order to compare our system, CheXNet \cite {rajpurkar2017chexnet} has been selected as our reference model. CheXNet is a CNN model trained to classify chest X-rays over 14 different lung diseases, \textcolor{Black}{which is a very similar problem} to ours (one of these lung diseases is precisely pneumonia). For this reason and the high performance with a similar problem was selected as the best model to compare the performance of ours.
In order to make an adequate comparison between both models, the number of neurons in CheXNet's last layer was changed from 14 (number of classes in the original paper) to two (number of classes in our work - consolidation / non-consolidation).
Once the final layer is modified, the rest of layers in CheXNet's are freezing except the last two, which we will re-trained using the target dataset}.
Taking a previously trained model in another similar domain and retraining it in the current dataset is a popular strategy in Deep Learning called \textit{transfer learning}.

\subsection{Performance metrics}

To analyze the performance of the different models, we used different metrics obtained from the ROC (Receiver Operating Characteristic Curve), a graphical representation that illustrates how the diagnostic capacity of a binary classifier system changes as its discrimination threshold is varied.
More specifically, this curve is created by plotting the true positive rate (TPR) against the false positive rate (FPR) at various threshold values \cite{fawcett2006introduction}.

A standard metric derived from the ROC is the \emph{Area Under the Curve} (AUC), which measures the overall quality and accuracy of the classifier. Finally, we also measured the TPR, which in our problem corresponds to the fraction of positive (consolidation) cases that are correctly detected by the model. \textcolor{Black}{Accordingly} \textcolor{Black}{to that, the fraction of patients that need antibiotics immediately that are correctly detected by the model}.

\subsection{Visual explanation using heatmaps}
\label{subsec:met_explainable_AI}

\textcolor{Black}{As it was previously mentioned, in order to generate an interpretable output for medical staff, we decided to generate a set of heatmaps. A heatmap is a matrix with the same size as the input image, where the value of each pixel is proportional to the importance of that pixel in the classification performed by the model has. Heatmaps were generated using the "Keras Vis" package \cite{raghakotkerasvis}. These are shown overlaid with the original X-ray to make them more interpretable by medical staff}. To overlay the images, the original X-ray and the obtained heatmap have a transparency degree of 50\%. This allows us to see under the heatmap.

Typically, binary classification problems use a single output neuron. However, we used two in order to obtain a separate heatmap and generate
the desired visualization for each of the two classes. Therefore our classifier layer has two output neurons, each one estimating the probability of the corresponding class (consolidation / non consolidation). The first step in heatmap generation is \textcolor{Black}{changing} "the activation function of the last layer of the network (the "output layer") from Softmax to Linear \cite{raghakotkerasvis}. \textcolor{Black}{These two heatmaps were generated to be able to compare both classifications, allowing the end-user to comparing the areas that the model considers relevant when determining if the patient presents consolidation, or non-consolidation.}

As explained above, ensembles are formed by five models. We generated ensemble heatmaps by averaging the individual heatmaps generated by those models. We also computed for each pixel the uncertainty of the heatmap by calculating the standard deviation of the individual heatmaps.

\section{Experimental results}
\label{experimental_results}
\subsection{CNNs trained from scratch versus transfer learning with CheXNet}
\label{subsect:own_models}

\textcolor{Black}{As described in \ref{met_network_architecture}, we considered six architectures for the CNN}. These architectures were compared using CheX-Net, which was retrained in our dataset using transfer learning (see \ref{subsection:reference-model}). Table \ref{tab:own_experiment_summary} shows the performance metrics of the different models. The statistics of AUC and TPR were calculated for all architectures using five different training / validation / test partitions.

\begin{table}[!h]
\centering
\begin{tabular}{|c||c|c|c|c|}
\hline
Arch   & \multicolumn{2}{c|}{AUC}            & \multicolumn{2}{c|}{TPR}            \\
         & Value      & p-value               & Value      & pvalue               \\
         \hline
\hline
\textbf{Arch 1}  & \textbf{0.80} $\pm$ \textbf{0.03} &                       & \textbf{0.62} $\pm$ \textbf{0.04} &                       \\
Arch 2  & 0.77$\pm$ 0.02  & \multirow{6}{*}{0.04} & 0.56$\pm$ 0.06
 & \multirow{6}{*}{0.80} \\
Arch 3  & 0.78$\pm$ 0.02 &                       & 0.55$\pm$ 0.08 &                       \\
Arch 4  & 0.76$\pm$ 0.02 &                       & 0.57$\pm$ 0.01 &                       \\
Arch 5  & 0.75$\pm$ 0.01 &                       & 0.55$\pm$ 0.07 &                       \\
Arch 6  & 0.77$\pm$ 0.02 &                       & 0.55$\pm$ 0.09 &\\
\hline
CheXNet & 0.76$\pm$ 0.02 &                       & 0.43$\pm$ 0.08        &                       \\ \hline
\end{tabular}
\caption{AUC and TPR values of our six architectures and CheXNet. Possible significant differences are analyzed using Anova One Way test ($\alpha = 0.05$).
}
\label{tab:own_experiment_summary}
\end{table}

\begin{figure}[h]
 \centering
 \includegraphics[width=3.50in]{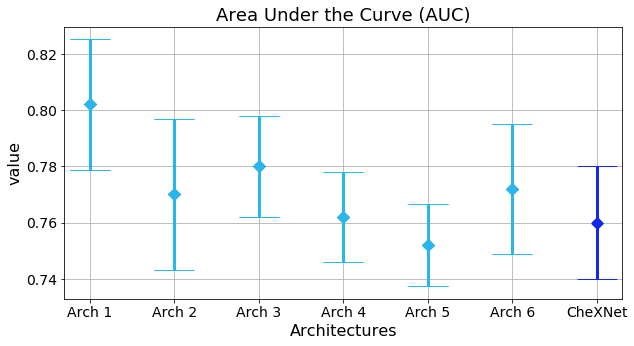}
   \caption{AUC values for CheXNet-based model (dark blue) and the CNN models trained from scratch.}
  \label{fig:auc_models}
\end{figure}

\begin{figure}[h]
 \centering
 \includegraphics[width=3.50in]{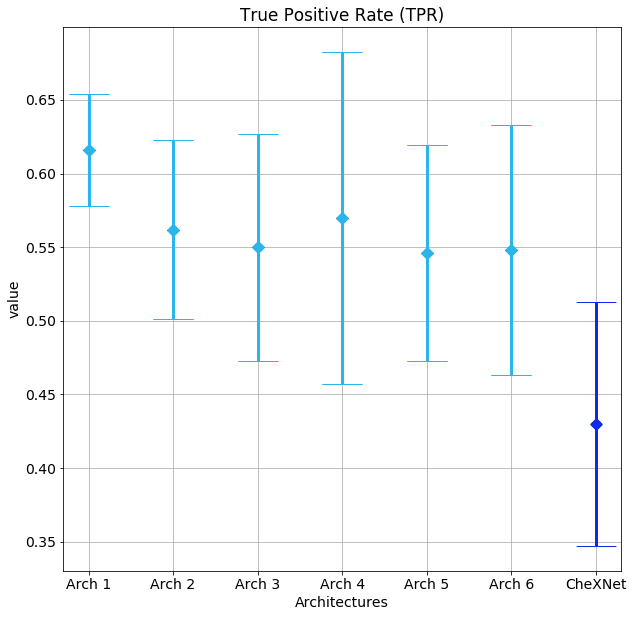}
   \caption{TPR values for CheXNet-based model (dark blue) and the CNN models trained from scratch.}
  \label{fig:tpr_models}
\end{figure}

\textcolor{Black}{Firstly, we can observe that our reference model, CheXNet, obtains a similar AUC value when is compared against our CNN architectures trained from scratch (see Figure \ref{fig:auc_models})}. However, the TPR obtained by CheXNet is 0.43 (Figure \ref{fig:tpr_models}). \textcolor{Black}{The reason might rely on the fact that CheXNet has been pre-trained using adult X-rays, and the differences between the original diseases are greater than the differences between alveolar and non-alveolar pneumonia. Furthermore, we can observe that the best architecture is Arch 1 (Architecture 1),} which achieves the highest AUC and TPR (see Table \ref{tab:own_experiment_summary}). Since the differences in AUC are statistically significant (p-value = 0.04), Arch 1 was selected as the best architecture.

In order to analyze Arch 1 in depth, we generated five different divisions of the dataset into construction / test subsets, and for each construction subset we generated five different training / validation splits. For each training / validation / test partition a model was trained from scratch using Arch1, \textcolor{Black}{obtaining a total of 25 models}. This allowed us to analyze the robustness and statistics of the architecture across different partitions (see Table \ref{tab:model2}).

\begin{table}[h]
\centering
\begin{tabular}{|c||c|c|c|c|}
\hline
Partition   & \multicolumn{2}{c|}{AUC}                              & \multicolumn{2}{c|}{TPR}                              \\
         & Value                        & p-value               & Value                        & p-value               \\ \hline \hline
1   & 0.78 $\pm$ 0.01 & \multirow{5}{*}{0.20} & 0.60 $\pm$ 0.05 & \multirow{5}{*}{0.44} \\
\textbf{2}   & \textbf{0.81} $\pm$ \textbf{0.01} &                       &\textbf{0.67} $\pm$ \textbf{0.06} &                       \\
\textbf{3}   & \textbf{0.80} $\pm$ \textbf{0.02} &                       & \textbf{0.62} $\pm$ \textbf{0.07} &                       \\
4   & 0.79 $\pm$ 0.01 &                       & 0.64 $\pm$ 0.10 &                       \\
5   & 0.80 $\pm$ 0.02 &                       & 0.71 $\pm$ 0.09 &  \\ \hline
Average   & 0.80 $\pm$ 0.01 &                       & 0.65 $\pm$ 0.04 &  \\ \hline
\end{tabular}
\caption{AUC and TPR values for Arch1 across the five different construction/test partitions. For each of them, five different training/validation splits were generated. Total number of CNNs trained from scratch: 25}
\label{tab:model2}
\end{table}

\begin{figure}[h]
 \centering
 \includegraphics[width=3.50in]{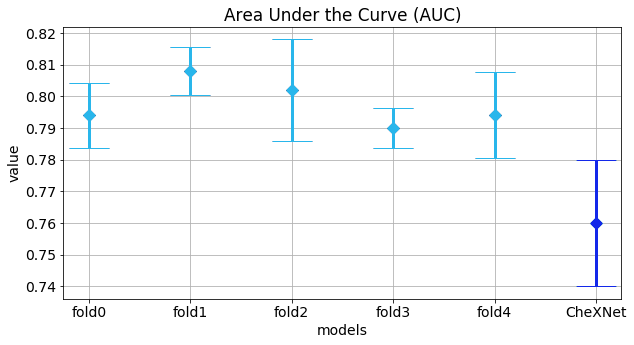}
   \caption{AUC for CheXNet and our different models.}
  \label{fig:auc_folds}
\end{figure}

\textcolor{Black}{We can observe in Table \ref{tab:model2} and Figure \ref{fig:auc_folds}} that the performance of the architecture across different construction / test partitions is very similar, and the variance of the metrics is low. Therefore, we can conclude that this architecture is robust over our dataset.

\textcolor{Black}{Table \ref{tab:model2} shows the performance of the individual Arch1 models}. We will now analyze the performance of an ensemble consisting of five individual models. \textcolor{Black}{Taking in consideration that for each construction/test partition we generated 5 different training/validation partitions and trained a different CNN from scratch for each of them. Now we will analyze the performance of an ensemble formed by those five individual models}. \textcolor{Black}{As it is shown in Table \ref{tab:ensembles1}, a clear improvement in performance, both in AUC and TPR values, is now obtained}. These values are higher than in the individual models with a difference of 9\% / 7\% for AUC/TPR respectively. \textcolor{Black}{Therefore, it can be concluded} that the ensemble is more robust against overfitting and provides better generalization capacity than the individual CNNs.

\begin{figure}[H]
 \centering
 \includegraphics[width=3.50in]{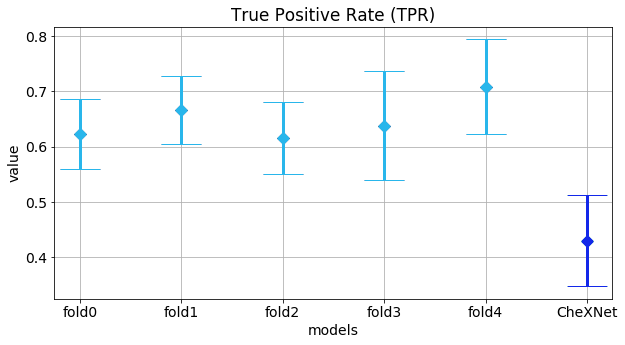}
  \caption{TPR values for the five models and CheXNet.}
  \label{fig:tpr_folds}
\end{figure}

\begin{table}[H]
\centering
\begin{tabular}{|c||c|c|}
\hline
Partition & AUC & TPR \\ \hline \hline
1 & 0.89 & 0.71 \\
2 & 0.92 & 0.73 \\
3 & 0.88 & 0.65 \\
4 & 0.88 & 0.73 \\
5 & 0.87 & 0.79 \\ \hline
Average & \textbf{0.89} $\pm$ \textbf{0.02} & \textbf{0.72} $\pm$ \textbf{0.04} \\ \hline
\end{tabular}
\caption{AUC and TPR values for Arch 1 ensembles.}
\label{tab:ensembles1}
\end{table}

\subsection{Visual explanation using heatmaps}
\label{subsect:heatmaps}
The last step in our system is the heatmaps generation.

\subsubsection{Individual heatmaps}

Figure 6 shows the prediction and heatmaps of an individual CNN for a non-consolidation test sample. The non-consolidation probability estimated by the model is 99.9\%. \textcolor{Black}{If we analyse and compare the heatmaps, we can see that neuron 0 heatmap (non-consolidation) is clearly brighter than neuron 1 heatmap (consolidation)}, and in neuron 1 heatmaps there are different areas marked that shouldn't be, corresponding to the clavicle and diaphragm. The conclusion is that the CNN has not found any signs of consolidation, but the previously referred areas of neuron 1 heatmap should not be lighted up.

\begin{figure}[H]
\centering{}
    \mbox{\includegraphics[width=4in, scale=0.4]{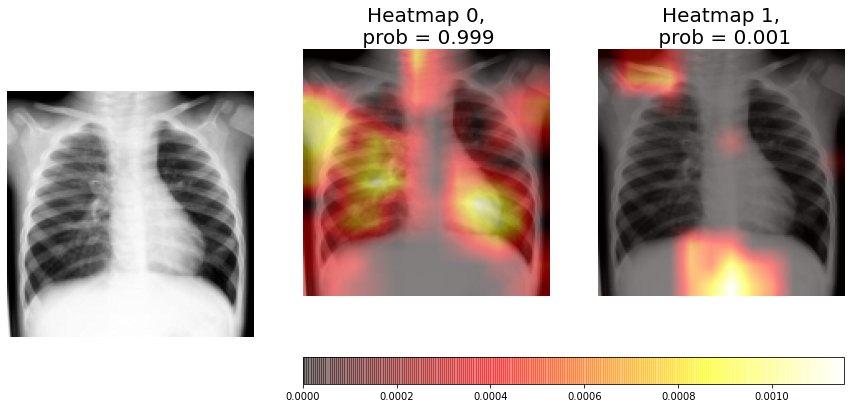}}
   \caption{Heatmaps generated with an individual CNN (Arch1) for a \textbf{non-consolidation} test X-ray (sample 1).
  \label{fig:heatmap_non_consolidation_simplified}}
\end{figure}

\begin{figure}[H]
  \centerline{
    \mbox{\includegraphics[width=4in, scale=0.45]{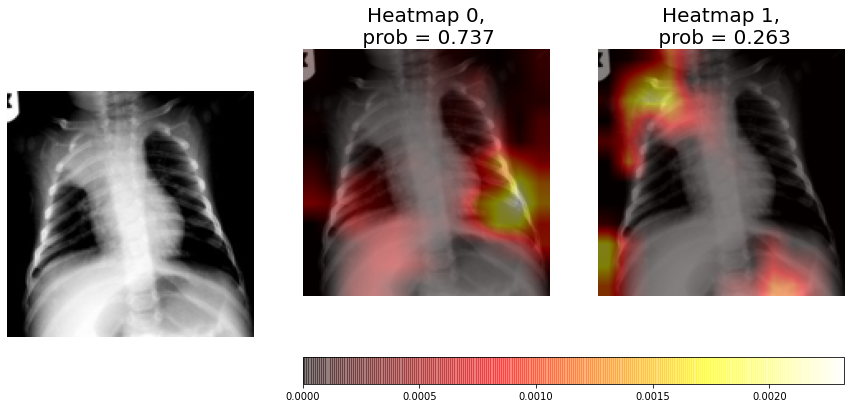}}
  }
  \caption{Heatmaps generated with an individual CNN (Arch1) for a \textbf{consolidation} test radiography (sample 1).}
  \label{fig:heat_consolidation_0_simplified}
\end{figure}

\begin{figure}[H]
  \centerline{
    \mbox{\includegraphics[width=4in, scale=0.45]{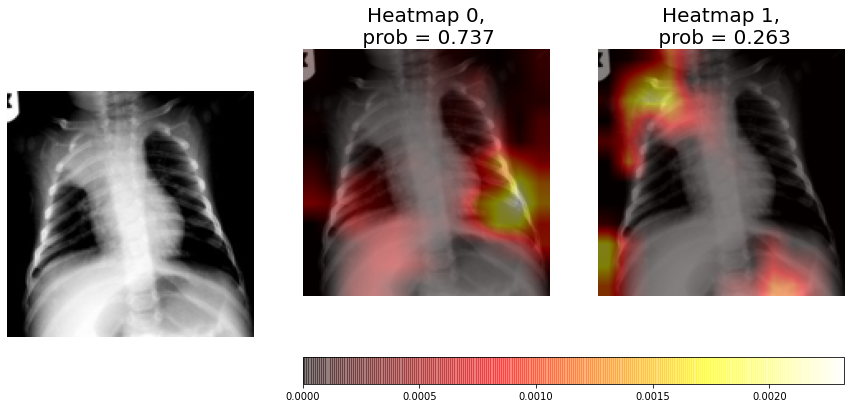}}
  }
  \caption{Heatmaps generated with an individual CNN (Arch1) for a \textbf{consolidation} test radiography (sample 1).}
  \label{fig:heat_consolidation_0_simplified}
\end{figure}

In Figure \ref{fig:heat_consolidation_0_simplified} we can see in neuron 1 heatmap that there is a consolidation sign in the upper left lung, but the consolidation probability estimated by the model is only 26.3\%. The conclusions is that the CNN model incorrectly classified the X-ray as a non-consolidation sample but the heatmap is able to find the consolidation evidences.

Figure 7 shows that there are signs of consolidation in the left lung. We can observe how both the probability estimated by the CNN model, 98.7\%, and the heatmap corresponding to the consolidation class show that the system has correctly classified the radiography. However, we can see how the heatmap only marks the left side of the consolidations area, which does not cover all the consolidation signs. We can conclude that the CNN model has correctly classified the sample but it has not found all interesting area of the image.

\subsubsection{Ensemble heatmaps}

\textcolor{Black}{The ensembles are composed by five individual CNNs, therefore, for each radiography we have five different heatmaps for each class (consolidation / non-consolidation)}. This allows us to compute the average heatmap and the uncertainty at each pixel (standard deviation at each pixel) for each class.

\begin{figure}[H]
  \centerline{
    \mbox{\includegraphics[width=4in, scale=0.45]{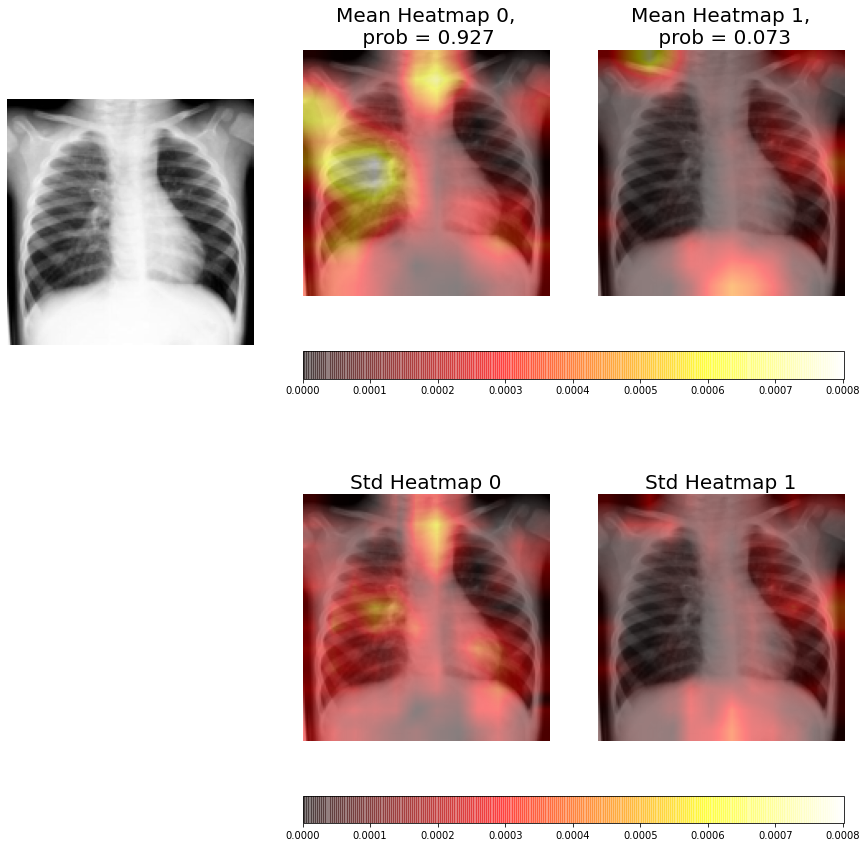}}
  }
  \caption{Heatmaps of a \textbf{non-consolidation} sample 1. Left side, Neuron 0, shows the heatmaps for the non-consolidation class, whereas the right side, Neuron 1, shows the heatmaps  for consolidation class for five individual CNNs.}
  \label{fig:heatmap_non_consolidation_complete}
\end{figure}

\begin{figure}[H]
  \centerline{
    \mbox{\includegraphics[width=4in, scale=0.45]{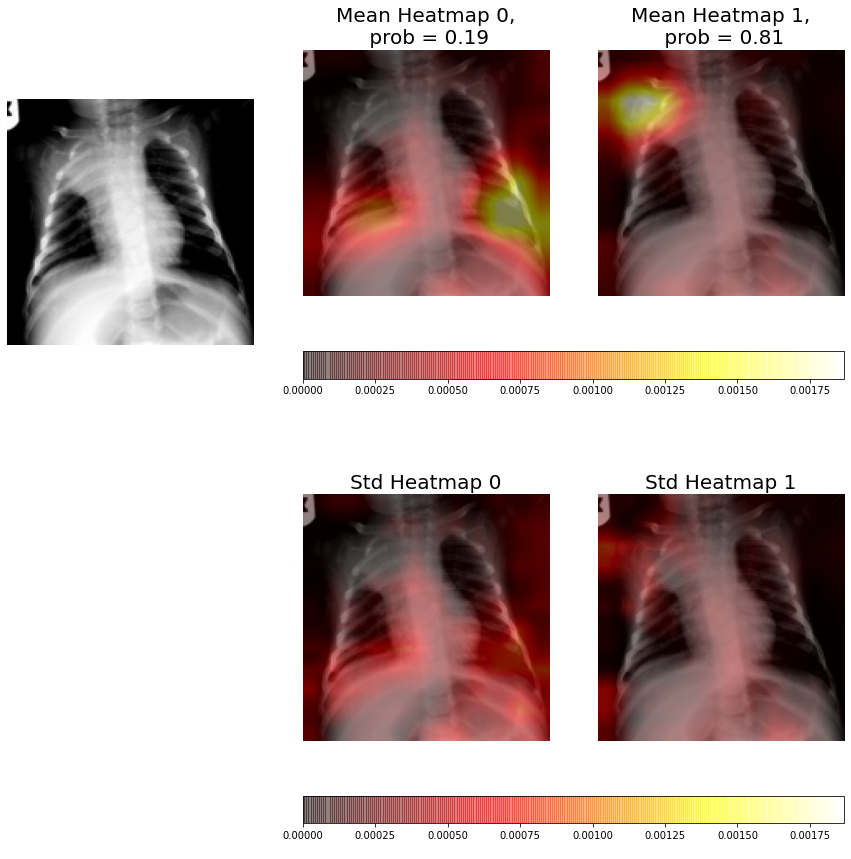}}
  }
  \caption{Heatmaps of a \textbf{consolidation} sample 1. Left side, Neuron 0, shows the heatmaps for the non-consolidation class, whereas the right side, Neuron 1, shows the heatmaps  for consolidation class for five individual CNNs.}
  \label{fig:heatmap_consolidation_complete_0}
\end{figure}

\begin{figure}[H]
  \centerline{
    \mbox{\includegraphics[width=3.7in, scale=0.45]{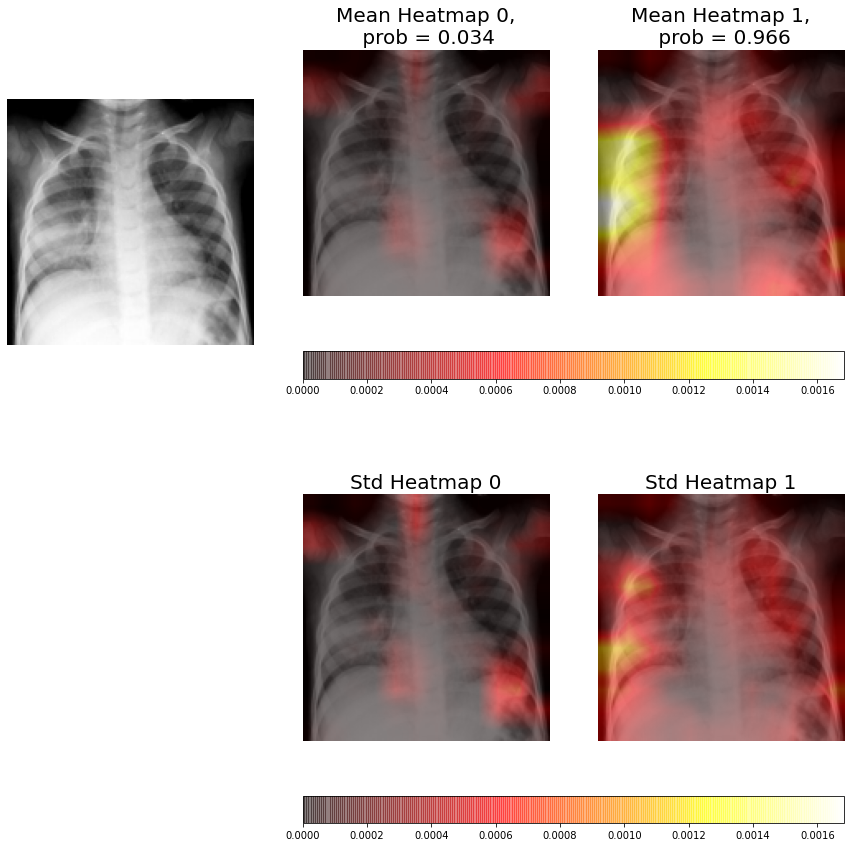}}
  }
  \caption{Heatmaps of a \textbf{consolidation} sample 2. Left side, Neuron 0, shows the heatmaps for the non-consolidation class, whereas the right side, Neuron 1, shows the heatmaps  for consolidation class for five individual CNNs.}
  \label{fig:heatmap_consolidation_complete_1}
\end{figure}

In Figures \ref{fig:heatmap_non_consolidation_complete}-\ref{fig:heatmap_consolidation_complete_1} we can see from left to right, and from top to bottom, the original X-ray, the average heatmap for the "non-consolidation" class, the average heatmap for the "consolidation" class, and the standard deviation heatmaps.

\textcolor{Black}{This visual representation provides to medical staff with relevant information. Firstly, the probabilities of each class predicted by the model are shown in the title of each average heatmap.
Secondly, the average heatmaps} show the areas of the X-rays that are most informative according to the ensemble.
Finally, the standard deviation heatmaps provide information about the areas of greatest disagreement among individual CNNs (that is, the areas with the greatest uncertainty).
This suggests that medical staff should pay more attention to areas with higher values in the average heatmaps and in the standard deviation heatmaps.

In Figure \ref{fig:heatmap_non_consolidation_complete} we can see that the average heatmap for neuron 0 (non-consolidation class) is much brighter than the average heatmap for neuron 1 (consolidation). If we compare this visualization with the \textcolor{Black}{one} obtained by an individual CNN (Figure \ref{fig:heatmap_non_consolidation_simplified}), we can see that it analyzes left lung in greater depth than the stand alone model, and focuses on the most relevant features to perform the X-ray classification. It also has a low standard deviation in this area and, unlike the visualization obtained by the individual CNN, it is hardly fixed on the heart area. \textcolor{Black}{Therefore, it can be concluded that the ensemble result is better than the one obtained by the individual CNN.}

\textcolor{Black}{In Figure \ref{fig:heatmap_consolidation_complete_0} we can see how the average heatmaps of both neurons have lighted up, whereas only neuron 1 heatmap should light up because signs of consolidation appears in the upper area of the left lung.} We can see how the standard deviation heatmap of neuron 0 presents higher values than neuron 1, which means that results of neuron 1 (consolidation class) are more robust than those of neuron 0 (non-consolidation class). If we look at the consolidation class average heatmap, the marked area corresponds to the pathology signs and it also marks a larger area than the same heatmap of the individual CNN (Figure 7). Furthermore, if we look at the estimated probability of the models (81\%) and we compare it with the one obtained with the individual CNN (26.7\%), we can see that ensembles give greater robustness to the system and avoid possible misclassifications obtained with the stand alone models. obtained with the stand alone models.

Figure \ref{fig:heatmap_consolidation_complete_1} shows that the X-ray has consolidation signs in the left lung. The average heatmap marks practically all area affected by the pathology, unlike the visualization of an individual CNN that only marks a small part of the affected area. We can also observe \textcolor{Black}{how} the standard deviation heatmap for neuron 1 indicates the \textcolor{Black}{adjacent areas} to the one of interest. The probability of this class (consolidation) is 96.6\%, lower than the obtained by individual CNN (98.7\%), this is because the result is obtained from the average of the five models and not all of them have to present the optimal results, however by obtaining the average of all the models we increase the robustness of the model.

The average and standard deviation heatmaps (Figures \ref{fig:heatmap_non_consolidation_complete}-\ref{fig:heatmap_consolidation_complete_1}) provide very interesting and relevant information related to the ensemble output and how the system made a particular decision.
However, the main problem related to this approach is the computational and time-consuming requirements.
In our computer system, it took about 400 seconds to calculate the heatmaps for the five models in the ensemble.


\subsection{Results with Kermany et al. dataset}

\textcolor{Black}{Finally, we want to investigate the robustness of our approach now using the dataset of Kermany et al. \cite{kermany2018identifying}.}
Analogous to that paper, two classification problems were considered, normal versus pneumonia and bacterial versus viral pneumonia.
\textcolor{Black}{We used the same training/test sets partition as in that work,} and randomly subdivided training set into training (70\%) + validation (30\%) partitions. We trained 5 CNN models, each with a different training / validation partition. \textcolor{Black}{We considered Arch1 architecture for the individual models and followed a similar approach to previous experiments when this new dataset is considered.}

\textcolor{Black}{Table \ref{tab:comparison_kermany1} shows the results related to normal versus pneumonia classification problem}. We observe that individual Arch1 models achieve AUCs similar to that reported in \cite{kermany2018identifying}. On the other hand, the TPR of the individual models is smaller (0.91 versus 0.932). However, the ensemble formed by our individual models achieves an AUC of 0.976 and a TPR of 1.0. Therefore, the ensemble shows better robustness and metrics than our individual models and the model presented in \cite{kermany2018identifying} based on DenseNet and transfer learning.

\textcolor{Black}{Note that the metrics are better than our previous dataset (see section \ref{subsect:own_models}), which could be due} Kermany et al. dataset is larger than ours (5,856 X-rays versus 950) and the original images are higher quality (1-2 million pixels versus 200K), so the training set contains much more information for constructing the models.

\begin{table}[h]
\begin{tabular}{|l||c|c|}
\hline
                                                                             & AUC         & TPR       \\ \hline
Kermany et al. model & 0.968        & 0.932     \\ \hline
\begin{tabular}[c]{@{}l@{}}Individual Arch 1 models\end{tabular} & 0.966 $\pm$ 0.007 & 0.91 $\pm$ 0.02 \\ \hline
\begin{tabular}[c]{@{}l@{}}Arch 1 ensemble\end{tabular} & 0.976 & 1 \\ \hline
\end{tabular}
\caption{Kermany et al. dataset, normal versus pneumonia classification: comparison of AUC and TPR values originally reported by Kermany et al. to results obtained by our models.}
\label{tab:comparison_kermany1}
\end{table}

\textcolor{Black}{Table \ref{tab:comparison_kermany2} indicates} the results for the bacterial versus viral pneumonia classification problem. \textcolor{Black}{Again, it can be observed that individual Arch1 models achieve AUCs values similar to that reported in \cite{kermany2018identifying}}. On the other hand, the ensemble formed by the individual models achieves an AUC value of 0.964 \cite{kermany2018identifying}.

These results are remarkable as we obtained better AUCs using simple CNN ensembles than very deep CNN transfer learning techniques \textcolor{Black}{such as those shown in \cite{kermany2018identifying}}, where transfer learning with a DenseNet architecture with 121 convolutional layers was used.

\begin{table}[h]
\begin{tabular}{|l||c|c|}
\hline
                                                                             & AUC         & TPR       \\ \hline
Kermany et al. model & 0.940        & 0.886     \\ \hline
\begin{tabular}[c]{@{}l@{}}Individual Arch 1 models\end{tabular} & 0.94 $\pm$ 0.02 & 0.78 $\pm$ 0.05 \\ \hline
\begin{tabular}[c]{@{}l@{}}Arch 1 ensemble\end{tabular} & 0.964 & 0.791 \\ \hline
\end{tabular}
\caption{Kermany et al. dataset, classification bacterial versus virical: comparison of AUC and TPR values originally reported by Kermany et al. to results obtained by our models.}
\label{tab:comparison_kermany2}
\end{table}

\textcolor{Black}{We can see that the application of ensembles in the Kermany et al. dataset also presents an improvement} in the result for the bacterial and viral pneumonia classification obtained in AUC, with a difference of 2\%.
However, it can be observed that the improvement obtained is lower than that obtained in our dataset. The difference between the datasets is probably due to their quality. \textcolor{Black}{The dataset provided by Kermany et al. presents greater size and quality,} so the models can achieve better values using only individual models and they can \textcolor{Black}{adequately generalize}, while our dataset, as it is more limited, the individual models are not able to obtain the best possible results.

\section{Conclusions}
\label{discusions_conclusions}
\textcolor{Black}{A large number of Convolutional Neural Network architectures have been recently proposed to help and support medical staff in the pneumonia diagnosis task}.
In this work, \textcolor{Black}{a new Machine Learning system based on ensembles}, which combines XAI techniques and CNN models, has been designed for the childhood pneumonia diagnosis.
\textcolor{Black}{When a simple model, over the target dataset, is used an AUC of 0.81 and a TPR of 0.67 values are obtained. However, applying ensemble techniques the performance of the model is improved to an AUC of 0.92 and a TPR of 0.73 values for this ensemble model}. These results are in the line of the current state of the art results, although in some cases (as was described in Section \ref{related_work_evolution_techniques}) are lower. However, and opposite to other previous approaches from the state of the art, our CNN models classify between the presence or the absence of consolidation. The objective of this work was to analyze and study the applicability of XAI techniques in this domain, and applying our approach in a particular dataset (950 X-rays). Three main problems have been overcome, \textcolor{Black}{the small size of samples in our dataset (less than 1,000)}, the low quality of images, and the anatomical variability existing within the age range of our dataset \textcolor{Black}{(between one month and 16 years old)} \cite{andronikou2009advances},  \cite{lander2013paediatric}.

Related to the visualizations provided (heatmaps) as the main XAI technique used, they present adequate results and provide much more information to medical staff than other methods. However, as it was described before, the quality of the model will depends on the quality of data, for this reason in some cases, as shown in Figure \ref{fig:heat_consolidation_0_simplified}, the model doesn't work correctly highlighting areas outside the lungs. Therefore, the model could be improved by using other techniques, such as segmentation methods to focus only on the lungs, and increasing the quality of the dataset. Finally we compare the results of models trained with our dataset and the dataset published by Kermany et al. and the results show that we can obtain similar values without very deep convolutional neural networks or transfer learning techniques.

\section{Future work}

\textcolor{Black}{
Firstly, we would like to explore different preprocessing techniques, such as lung segmentation to force the model to focus only on the lungs, or different data augmentation techniques to improve the quality of the dataset. These techniques would improve the performance of our model.}
\textcolor{Black}{
Secondly, we will study the performance of this technique with other datasets: for example with different thoracic diseases, such as COVID-19, Covid Data Save Lives (from HM hospitals) \cite{hmhospitales}, that contains different kinds of clinical images and clinical data, BIMCV-COVID19 \cite{bimcv}, which it is composed of different kinds of medical imaging and testing COVID-19; or multi-label datasets, for example such as PadChest \cite{padchest}, composed by different clinical images of lung and heart diseases or COVID-19 Image Data Collection \cite{cohen2020covid}, that contains X-rays by five lung diseases. Also we would to compare our system with other architectures, such as: COVID-Net \cite{covidnet}, the model used by kermany et al. \cite{kermany2018identifying}, Multichannel Convolutional Neural Network \cite{nahid2020novel}  among others, to better understand and study the robustness of the approach presented.}

\textcolor{Black}{Other complementary XAI techniques based on visualizations will be considered with the aim to facilitate the work of medical staff. Finally, and to test the generality of the method proposed in this work, other domains, such as those related to industrial domains \cite{rodriguez2020conformance}, where the combinations of CNN ensemble models and XAI methods, could increase the capabilities (analysis, prediction, explainaibility) of current used methods, will be explored and tested in the near future.}

\section*{Acknowledgements}
    This work has been supported by Spanish Ministry of Science and Education under TIN2017-85727-C4-3-P (DeepBio) grant, by  Agencia Estatal de Investigación AEI/FEDER Spain, Project PGC2018-095895-B-I00, by CHIST-ERA 2017 BDSI PACMEL Project (PCI2019-103623), and by Comunidad Aut\'{o}-noma de Madrid under S2018/ TCS-4566 (CYNAMON),  S2017/BMD-3688 grants. We gratefully acknowledge the support of NVIDIA Corporation with the donation of the Titan V GPU used for this research.

\bibliographystyle{unsrt}  

\bibliography{biblio}

\end{document}